\documentclass[runningheads]{llncs}
\usepackage[T1]{fontenc}
\usepackage{graphicx}
\begin{document}
\title{A DataOps Toolbox Enabling Continuous Semantic Integration of Devices for Edge-Cloud AI Applications}
%A DataOps Toolbox Enabling Continuous Semantic Integration of Devices for Edge-Cloud Swarm Applications
%
\titlerunning{A DataOps Toolbox Enabling Continuous Semantic Integration}

% If the paper title is too long for the running head, you can set
% an abbreviated paper title here
%
\author{
Mario Scrocca\inst{1}\orcidID{0000-0002-8235-7331} \and
Marco Grassi\inst{1}\orcidID{0000-0003-3139-3049} \and
Alessio Carenini\inst{1}\orcidID{0000-0003-1948-807X} \and
Darko Anicic\inst{2}\orcidID{0000-0002-0583-4376} \and
Jean-Paul Calbimonte\inst{3,4}\orcidID{0000-0002-0364-6945} \and
Irene Celino\inst{1}\orcidID{0000-0001-9962-7193}
}
\authorrunning{M. Scrocca et al.}
% First names are abbreviated in the running head.
% If there are more than two authors, 'et al.' is used.
%
\institute{Cefriel -- Politecnico di Milano, Milan, Italy
\email{\{mario.scrocca,marco.grassi,alessio.carenini,irene.celino\}@cefriel.com}\\ \and
Siemens AG, Munich, Germany\\
\email{darko.anicic@siemens.com}\\ \and
HES-SO University of Applied Sciences and Arts Western Switzerland, \\Sierre, Switzerland \\
\email{jean-paul.calbimonte@hevs.ch} \\ \and
The Sense Innovation and Research Center, Lausanne, Switzerland}

\maketitle              % typeset the header of the contribution
\begin{abstract}
The implementation of AI-based applications in complex environments often requires the collaboration of several devices spanning from edge to cloud. Identifying the required devices and configuring them to collaborate is a challenge relevant to different scenarios, like industrial shopfloors, road infrastructures, and healthcare therapies. We discuss the design and implementation of a DataOps toolbox leveraging Semantic Web technologies and a low-code mechanism to address heterogeneous data interoperability requirements in the development of such applications. 
The toolbox supports a continuous semantic integration approach to tackle various types of devices, data formats, and semantics, as well as different communication interfaces.
The paper presents the application of the toolbox to three use cases from different domains, the DataOps pipelines implemented, and how they guarantee interoperability of static nodes' information and runtime data exchanges. Finally, we discuss the results from the piloting activities in the use cases and the lessons learned.

\end{abstract}
\section{Introduction}\label{sec:introduction}

The Internet of Things (IoT), combined with robotic agents, edge intelligence and cloud computational power, can offer numerous advantages across various sectors through the implementation of AI-based applications~\cite{disclosing2023barbuto}. To achieve this, collaboration and seamless data exchanges between devices are needed to empower functionalities that can enhance efficiency and productivity. IoT devices generate vast amounts of data, which, when analyzed, can provide valuable insights and improve decision-making processes. However, the realization of these benefits depends on the assumption that the data produced can be effectively utilized to empower intelligent applications. This is often not the case, as identifying the right set of devices in possibly dynamic environments and integrating their data can be quite challenging. Devices vary widely in capabilities, communicate through different protocols, and share information in various formats. Additionally, they may evolve or change their status over time, complicating data integration for application developers.

The objective of the SmartEdge\footnote{https://www.smart-edge.eu/} project is to support the low-code programmability of AI applications on the edge. The concept of Continuous Semantic Integration (CSI) is a key element of the proposed solution and leverages Semantic Web technologies to facilitate the configuration and execution of applications through interoperable semantic representations of devices, data, and applications.
First, device capabilities and application requirements are described using standardized vocabularies. This enables the semantic discovery of device capabilities and their alignment with application requirements. Additionally, the CSI ensures interoperability in data exchanges between devices through uniform and standardized interfaces. Once device capabilities are described and their data is integrated, the CSI facilitates the development of applications. Devices can be discovered, matched with requirements, and assembled into application workflows using the concept of low-code programming.
In this paper, we focus on the DataOps toolbox as an enabler of CSI to support data operations for the interoperability of static nodes' information and runtime data exchanges. The toolbox provides a set of modular and configurable blocks that can be combined and configured to address heterogeneous requirements in DataOps pipelines deployable from cloud to edge. The development and testing of such pipelines across industrial, automotive and health domains is discussed to demonstrate the solution's flexibility and effectiveness in addressing the existing challenges.

The remainder of this paper is organised as follows. Section \ref{sec:related} discusses related work applying Semantic Web technology to facilitate the development of edge-cloud AI applications. Section \ref{sec:challenges} introduces the use cases considered and the associated challenges. Section \ref{sec:csi} describes the CSI approach envisioned and the need for a DataOps solution. Section \ref{sec:dataops} outlines the requirements for the DataOps toolbox, its design and implementation. Section \ref{sec:usecases} explains how the toolbox is effectively adopted to address each use case. Section \ref{sec:validation} evaluates the advantages and limitations of the proposed solution, highlighting the lessons learned in its application to practical scenarios. Finally, Section \ref{sec:conclusions} draws the conclusions and presents the future work.

\section{Related Work}\label{sec:related}

Semantic Web technologies have increasingly been applied to the Internet of Things (IoT) to address challenges such as heterogeneity, interoperability, and scalable data integration.

Industrial IoT (IIoT) environments benefit from semantic descriptions of devices, services, and data to enable dynamic discovery, orchestration, and reasoning \cite{amara2022}. Recent research has explored the application of semantic modeling directly within IoT development tools. For instance, Node-RED has been exploited as a visual tool for the rapid development of semantically interoperable IIoT applications, providing domain engineers with intuitive means to create semantic flows and annotate devices and data~\cite{Thuluva_2020}.
The challenge of semantic interoperability across heterogeneous industrial systems has been widely studied. The integration of OPC UA\footnote{\url{https://opcfoundation.org/}}, a de facto industrial standard, with Semantic Web technologies has gained traction as a strategy to enrich device communication with semantic meaning. Formal representations of OPC UA specifications through RDF and OWL~\cite{Bi_2024,Schiekofer_2019} allow for querying, enable reasoning, and facilitate interoperability across diverse device ecosystems.

Standardization efforts, such as the W3C’s Web of Things (WoT) recommendations, play a pivotal role in this landscape. Surveys such as~\cite{Sciullo_2022} provide a comprehensive overview of the current state of WoT and emphasize the role of semantic annotations in enabling automation and dynamic service composition. Furthermore, these recommendations can support knowledge graph construction from WoT data sources, as shown in~\cite{Assche_2021}. The WoT TD (Thing Description) specification~\cite{thingdescription2023}  supports the interoperable description of devices. The Domus TDD API\footnote{\url{https://github.com/eclipse-thingweb/domus-tdd-api}} implements the WoT Discovery specification to search and retrieve Things according to specific requirements. The Sensor, Observation, Sample, and Actuator (SOSA) ontology~\cite{Janowicz_2019} provides a lightweight vocabulary to describe a wide range of sensing applications with reduced modelling complexity.

Distributed edge computing systems also stand to benefit from semantic descriptions. In this direction, the authors in~\cite{Anuraj_2023} explore agent-based orchestration on edge swarms, leveraging semantics to support flexible deployment and runtime adaptation. A systematic review~\cite{Anuraj_2024} further consolidates these findings and underscores the emerging trends.

Semantic data conversion approaches have been proposed~\cite{Bennara_2020,scrocca2020turning} to guarantee interoperability among heterogeneous information systems via an any-to-one centralised
mapping approach that transforms data in two steps: from the input data format to the reference ontology (\emph{lifting}) and from the reference ontology to the target data format (\emph{lowering}). There are several possible implementations of the described conversion procedure, e.g., based on an
Object-Relational Mapping (ORM) approach~\cite{carenini2018st4rt} or based on declarative mappings for the lifting (e.g., the R2RML~\cite{r2rml} W3C recommendation for relational databases and
RML~\cite{iglesias2023rml} for heterogeneous data sources) and lowering phases (e.g., combining declarative data access with template
engines~\cite{scrocca2024not,scrocca2020turning}).The open-source Chimera framework~\cite{grassi_composable_2023} offers a modular and configurable approach to build flexible pipelines for data integration based on Semantic Web technologies.
The authors in~\cite{scrocca2024intelligent} discuss the application of the Chimera framework to enable interoperability of mobility data sources and support Intelligent Transportation Systems (ITS). Finally, the stream reasoning community is also exploring declarative approaches for the generation of RDF streams from heterogeneous data sources~\cite{grounding2024bonte,rmlstreamer2022}.

This paper discusses the DataOps toolbox that supports Continuous Semantic Integration to simplify developing and deploying intelligent swarm applications via Semantic Web technologies. Additionally, we focus on providing low-code tooling that could support non-expert users in applying this approach. The solution builds on the state-of-the-art presented and is applied to concrete use cases from different domains to evaluate their effectiveness, scalability and impact. Complementary topics are addressed in the context of SmartEdge to deliver a complete toolchain for low-code programmability of intelligent edge applications, leveraging the collaboration of distributed devices as a swarm. On the one hand, a specific focus is given on networking operations to optimise latency of data exchanges and support dynamic onboarding of devices within a swarm~\cite{smartedged42}. On the other hand, implementing an integrated approach for specific AI applications like multi-modal scene understanding is investigated together with methods to support dynamic balancing of edge and cloud resources within a swarm~\cite{smartedged52}.

\section{Use Cases}\label{sec:challenges}

This section outlines three relevant use cases from the SmartEdge project in diverse domains to highlight the target challenges for the DataOps toolbox.

\textbf{Low-Code Edge Intelligence for Smart Factory} The use case focuses on a manufacturing scenario where a production line should be configured to assemble order-specific products according to individual customer requirements. The demonstrator consists of three workstations: the assembly module for assembling the product based on user specifications, the mini-backbone module for transporting and distributing the product carrier, and the shipping module for delivering the product. The operator should be able to select the order in which different input components should be placed in the final product according to the custom order.
The objective is to enable a low-code interface for the insertion of this information, providing flexible means to reprogram the production line. The main challenge of this use case lies in enabling the discovery and collaboration of heterogeneous physical devices and software components running AI-based solutions (e.g., visual checking of the crafted product). Predominantly, industrial devices adopt the OPC UA standardized information models and interfaces. However, this use case also utilizes IoT devices, which are integrated through the WoT standard. Both device types need to be described in a structured and interoperable format to enable their low-code programmability into integrated applications.
The WoT standard's Thing Descriptions already adopt RDF and a common vocabulary for their representation. Moreover, WoT defines a mechanism for the discovery and retrieval of Thing Descriptions. On the contrary, the device descriptions in the OPC UA standard need a dedicated transformation and discovery process.
This overall approach accelerates shopfloor application development to meet ever-changing production requirements. The use case is implemented on a demonstrator in the Siemens manufacturing laboratory in Nuremberg, Germany.

\textbf{Smart Vehicle to Infrastructure} The use case deals with developing intelligent applications leveraging the road infrastructure and connected smart vehicles. The goal is to support, without manual programming, the dynamic computation of traffic indicators on specific areas for intelligent traffic management applications. This requires the combined processing of multi-modal (e.g., camera, lidar, radar, etc.) data streams from various devices (fixed and mobile) via AI-enabled algorithms for scene understanding. Different sensors are processed using various machine learning models, which may also be directly supplied and integrated by the sensor vendor and therefore cannot be modified. This causes heterogeneity in the generated outputs and limits the possibility of having an integrated real-time understanding of the traffic situation. Moreover, existing data sources, like real-time data from public transport authorities, can be combined to provide additional information on incoming vehicles. Based on the traffic indicators, all the signals within one road intersection may be used to execute intelligent traffic management algorithms, e.g., acting on the traffic light as an actuator and/or providing real-time alerts to connected vehicles. The main challenge in this use case deals with the harmonisation and fusion of heterogeneous data streams from multiple devices. Additionally, the computation should happen on the edge within Road Side Units (RSU) at each intersection to reduce the latency. For this reason, minimal resource consumption is required.
The use case is tested in the Mobility Lab in Helsinki and leverages instrumented road intersections and two vehicles equipped with sensors and capabilities to communicate with the infrastructure and other vehicles.

\textbf{Smart Health Rehabilitation} The use case focuses on human activity tracking for physiotherapy exercises, and more concretely on neck rehabilitation. 
These therapies are typically prescribed following a surgical intervention (post-sub-acute phase), injury recovery, or chronic conditions. 
As opposed to traditional treatment with only episodic verification of the therapist, the objective is to allow patients to perform prescribed exercises at home, supported by personalised assistive technologies. The envisioned solution entails the interoperability of heterogeneous sensor nodes (comprising wearables for biomarker monitoring and environmental sensors) and personal devices (e.g., a tablet) to execute and monitor exercises~\cite{BuzcuCAC24}. 
In these settings, the therapist can define the exercises through a visual interface and then the system automatically orchestrates the set of devices available for a specific patient accordingly. 
Exercises may include strength and flexibility conditioning programs (e.g., neck lateral and frontal movements, compensation of shoulder movement, etc.), which require data exchange and integration from different devices to correctly interpret the patient's movements in the physical space. 
The main challenge of this use case is related to the fact that each patient may be given a different set of devices, thus requiring the adaptation of the defined exercise to the current setting. The use case is being tested in collaboration with the PhysioLab of the School of Health at HES-SO in Leukerbad, Switzerland. 

\section{Continuous Semantic Integration}\label{sec:csi}

\begin{figure}[t!]
    \centering
    \includegraphics[width=\linewidth]{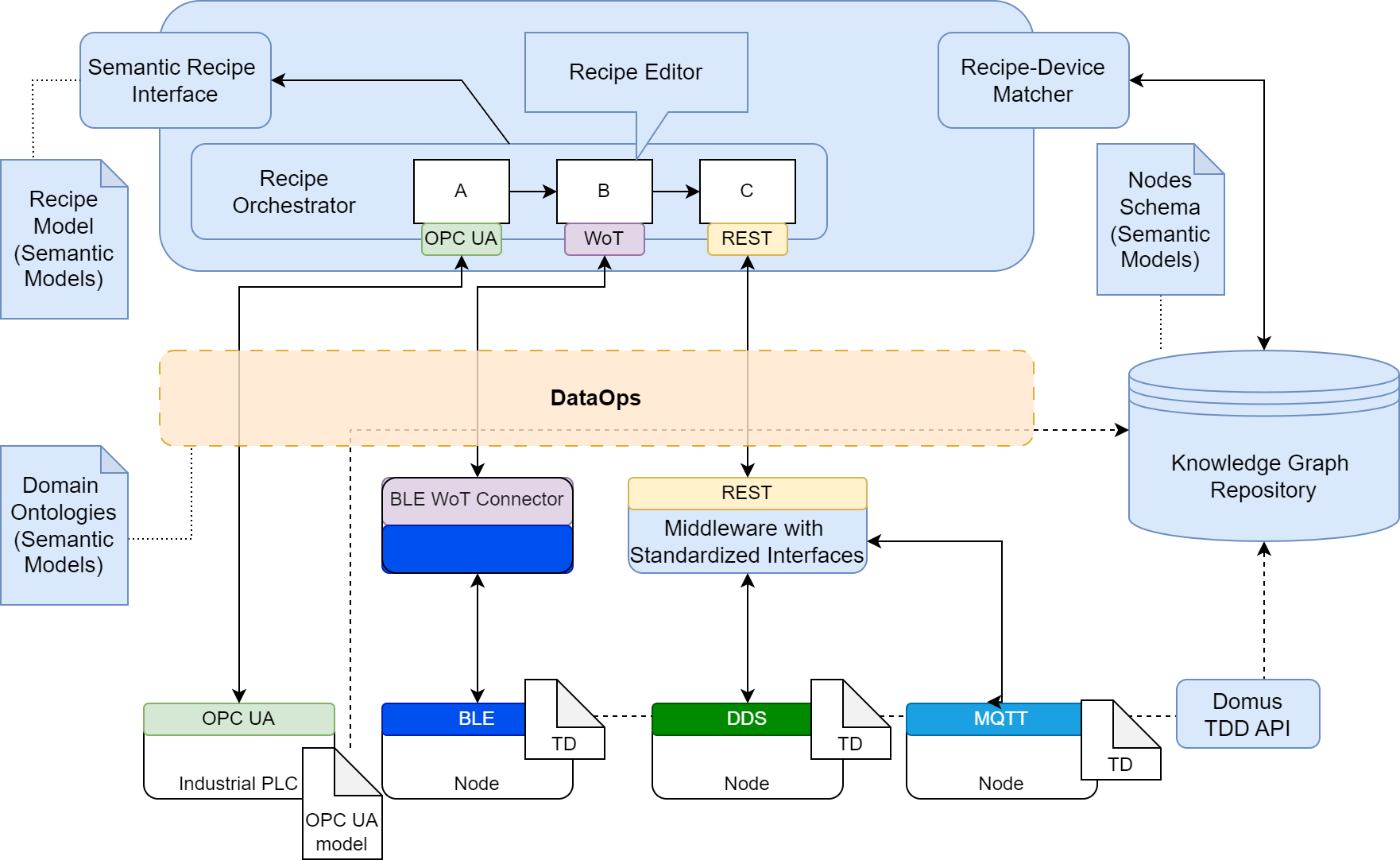}
    \caption{Continuous Semantic Integration overview diagram.}
    \label{fig:csi}
\end{figure}

This section provides an overview of the designed concept of Continuous Semantic Integration (CSI), outlining the need for DataOps operations. CSI aims to enable a low-code approach to define distributed intelligent applications. The diagram in Figure~\ref{fig:csi} shows the main components supporting this approach, and we discuss in the following paragraphs how they enable different tasks required for implementing CSI within a specific use case. We use the term \emph{node} to refer to a component that exposes a set of affordances supporting an AI-based application (e.g., produces or processes a data stream). A node may correspond to a device (e.g., a robot), or multiple nodes may be deployed on the same device. We use the term \emph{swarm} to refer to the group of nodes that collaborate to execute the application. The \emph{swarm} may adopt orchestration (centralized control coordinating tasks), choreography (decentralized interaction via events) or swarm intelligence (collective behaviour from simple agents).

\textbf{Interoperable description of nodes and their capabilities.} The first task is based on an interoperable description of the nodes available for a certain use case, i.e., the list of nodes that can be selected to form a swarm and execute an application. The description of the nodes should be compliant with a standard semantic model (i.e., the \emph{Nodes Schema}\footnote{\emph{SmartEdge Schema} available at \url{https://w3id.org/smartedge/smartedge-schema}}) and stored within a Knowledge Graph Repository (KGR). The KGR provides default support for W3C Thing Descriptions, leveraging the Domus TDD API, but can be extended to support additional semantic descriptions compliant with the \emph{Nodes Schema}.
DataOps operations can support the harmonisation of node descriptions and potential data integration requirements for their storage and retrieval from the KGR.

\textbf{Enable standardized communication interfaces for relevant
nodes.} The second task is associated with the need for enabling communications among nodes leveraging different protocols. Standard communication interfaces can be enabled with the support of a common middleware solution.
As an example, Zenoh\footnote{\url{https://zenoh.io/}} offers a suitable solution to expose MQTT and DDS streams as REST APIs. Alternatively, a WoT Serviant as defined in the WoT Architecture~\cite {wotarchitecture11} can be leveraged, e.g., for Bluetooth devices.

\textbf{Define mediated data exchanges ensuring semantic interoperability.} DataOps operations support mediated data exchanges among nodes in the swarm. They enable semantic conversion processes and guarantee not only the exchange of data among nodes, but also the required schema and data transformations for semantic interoperability, e.g., considering the semantics of a specific domain ontology.

\textbf{Support execution of swarm applications on cloud and edge.} The execution of DataOps operations can be configured according to the deployment environment (edge or cloud), the resource constraints of the nodes and the scalability needs of the application.

\textbf{Define interoperable recipes for swarm applications.} The \emph{Recipe Model}\footnote{\emph{SmartEdge Recipe Model} available at \url{https://w3id.org/smartedge/recipe-model}} is an ontology enabling the specification of an application template in terms of the capabilities required by a set of devices for its execution. The steps and the application logic to implement a recipe can be defined using a low-code approach, like the one enabled in SmartEdge by using the Mendix low-code tool\footnote{\url{https://www.mendix.com/}}. The \emph{Recipe Model} guarantees the interoperability of recipes defined by different low-code tools. Recipes can be stored and retrieved from the KGR, and once a suitable recipe is identified, the low-code developer can customise it and finalise the application design.

\textbf{Execute interoperable recipes on a swarm.} Given a recipe and the description of available nodes, a dedicated component is provided to match the required capabilities with the affordances of available nodes to identify a suitable swarm for execution. An orchestrator may be deployed at runtime to support the application execution among the nodes.

\section{DataOps Toolbox Design and Implementation}\label{sec:dataops}

The issue of data interoperability is a significant concern when
operating within a multi-stakeholder ecosystem comprising diverse actors. Similarly, within a swarm there is a need for data
interoperability among diverse nodes that employ heterogeneous data formats, specifications, and semantics. To enable the CSI approach presented, two scenarios should be covered: (i) \emph{interoperability of static information} describing the application, the nodes and their capabilities, and 
(ii) \emph{interoperability at runtime within a swarm} of dynamic nodes' information and data exchanges between nodes. 
The DataOps toolbox is designed to provide a flexible solution to address this two scenarios and cover the first two layers defined in the European Interoperability Framework (EIF)\footnote{\url{https://joinup.ec.europa.eu/collection/nifo-national-interoperability-framework-observatory/solution/eif-toolbox/6-interoperability-layers}}: (i) \emph{heterogeneous interface integration} to guarantee technical interoperability (i.e., the possibility of a data exchange between two systems) through standardized interfaces, and (ii) \emph{payload conversion} to guarantee semantic interoperability (i.e., ensure that the target node can understand the message received and act appropriately) through standardized semantics.

For the definition of more non-functional requirements, it is important to consider which type of data is processed by a pipeline and its needs in terms of performance and scalability.
On one hand, \emph{static data exchanges} often require the conversion of datasets of larger size with low frequency, thus requiring the minimisation of resource usage and scalability in terms of the input size. On the other hand, \emph{runtime data exchanges} are usually associated with small-size messages with high frequency, thus requiring the minimisation of the latency introduced by the conversion process and scalability in terms of concurrent conversion requests. An additional constraint is related to the types of nodes involved in the considered use cases. The resources available for each node, mainly CPUS and RAM, should be taken into account especially for edge devices.

Additionally, the DataOps toolbox should support deployments either in the cloud or on the edge, considering different types of devices (e.g., hardware and operating system) as well as cloud environments such as container orchestrators. To facilitate the reusability and portability of DataOps pipelines, their definition should be decoupled from the necessary configuration for the deployment. 
Finally, the toolbox should support low-code approaches to minimise the amount of code to be written and increase the reusability of already defined solutions.

\begin{figure}[t!]
    \centering
    \includegraphics[width=0.8\linewidth]{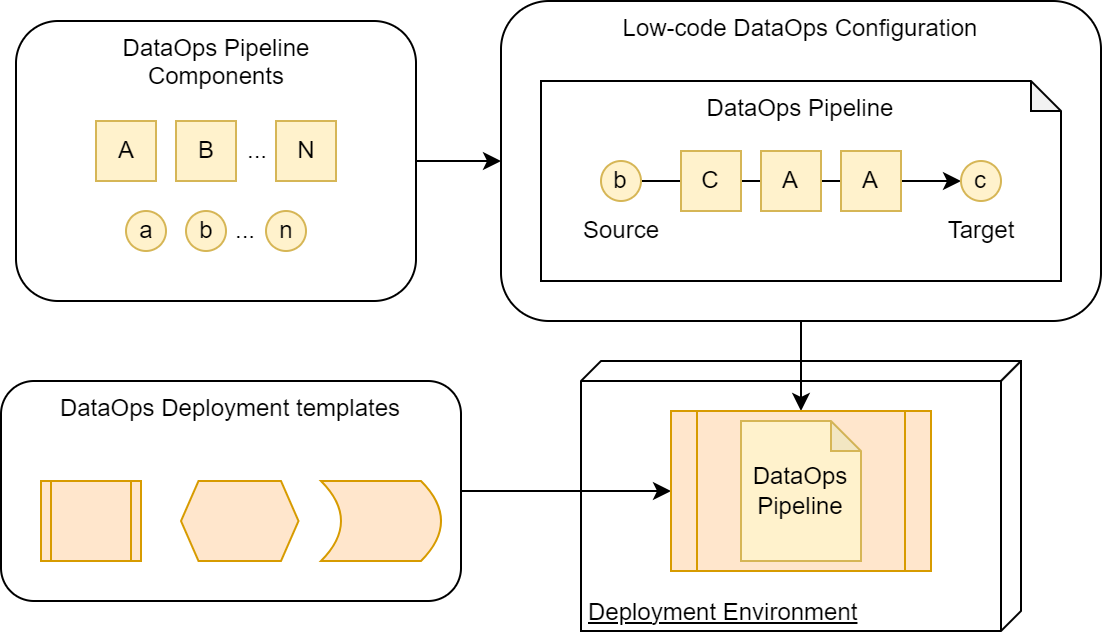}
    \caption{DataOps Toolbox overview diagram.}
    \label{fig:toolbox}
\end{figure}

Considering these requirements, the DataOps toolbox is designed as composed of three main artefacts as shown in Figure \ref{fig:toolbox}. The first artefact identifies the modular and configurable components made available to define a DataOps pipeline. The second artefact provides tools for the low-code definition of the pipelines. The third one offers a set of templates to enable the deployment of the same pipeline in different environments.

The DataOps toolbox is defined to support different deployment strategies: (i) within a dedicated node (mediation node), (ii) embedded in the swarm orchestrator (mediation service), (iii) embedded in the middleware, (iv) embedded in the source/target node. Each option is associated with different trade-offs and depends on the specifics of the considered scenario. In the first two cases, the orchestrator could possibly enable the deployment of a mediation node or the execution of a mediation service by considering the requirements of the application to be executed and the nodes executing it. The last two cases assume a predefined configuration for either the middleware or specific nodes to enable their cooperation in the application run by the orchestrator.

From the state-of-the-art analysis~\cite{smartedged31}, the declarative semantic conversion process is selected as the approach to enable data interoperability and data integration. Declarative mapping rules are leveraged to configure to/from transformations from a reference conceptual model relying on Semantic Web technologies for syntactic and semantic interoperability.
The mapping rules are decoupled from the component responsible for their execution to improve their maintainability and reusability.
To implement such an~approach, the main types of building blocks for a DataOps pipeline are the \emph{Node Connector}s, which are blocks responsible for enabling data exchanges with different types of interfaces/protocols, and the \emph{Mapping Processor}s, which are blocks capable of executing declarative mapping rules for data and schema transformations. Additional blocks may be integrated within a pipeline to perform additional manipulations or to implement \emph{Enterprise Integration Patterns} (EIP) \cite{hohpe2004enterprise}.

The DataOps toolbox is built relying on the Apache Camel\footnote{\url{https://camel.apache.org/}} framework as a solution enabling enterprise integrations by configuring building blocks within an executable pipeline. Moreover, Camel offers many production-ready components that can be easily integrated within a pipeline as \emph{Node Connector} for common protocols and interfaces. Finally, Camel supports EIPs and can be easily extended to define custom components to be integrated within a pipeline. As a reusable component for Camel, the \emph{Chimera} framework introduces operations for constructing, manipulating, and exploiting knowledge graphs within a pipeline. It provides support for operations on an RDF graph (either local or remote), execution of lifting and lowering operations. To support heterogeneous DataOps operations with different performance and scalability requirements, we redesigned the template-based mapping rule component (\texttt{mapping-template}) to enable knowledge conversion between different data representations and thus supporting both lifting and lowering~\cite{scrocca2024not}. The tool leverages a Mapping Template Language (MTL) that builds on existing approaches for declarative KG construction but trades off a fully declarative approach to provide more flexibility in the definition of the mapping rules.
We performed a complete refactor of the Chimera framework to increase the solution's overall TRL, improve its reusability and facilitate the configurability and composability of pipelines~\cite{grassi_composable_2023}. Finally, we developed the \texttt{typhon-rml} tool~\cite{grassi2025typhon} to support the automatic generation of a DataOps pipeline and MTL mapping rules from an RML mapping document. This approach facilitates the reusability and customisation of RML mapping rules across different integration scenarios.

To support different deployment options on edge devices and in the cloud, we investigated the alternative options for executing Camel integrations. We defined and implemented reusable templates\footnote{\url{https://github.com/cefriel/chimera-deployment-templates}} to facilitate the deployment of DataOps pipelines in different contexts. We enable alternatives spanning from integration as a library to container orchestration\footnote{A description of the characteristics of each alternative, and the analysis of the pros and cons, can be found in~\cite{smartedged32}}. An important aspect for edge deployments is the possibility of building directly executable binary files that minimise resource usage. On the contrary, execution in the cloud enables flexible scalability according to the current demand adopting scale-to-zero and autoscaling approaches.

Finally, we investigated how to facilitate the implementation of DataOps pipelines via low-code approaches. First, the adopted solution emphasises configuration over manual coding. The declarative configuration of components reduces the need to implement  solutions from scratch.
A set of abstractions empowers a low-code approach to data integration, as it exposes all available functionalities of DataOps components through well-documented URI parameters. Users can configure the pipelines
using one of the Camel domain-specific languages (DSL).

To streamline the definition of pipelines, we also developed support for the \emph{Karavan}\footnote{\url{https://github.com/apache/camel-karavan}} tool that provides a drag-and-drop user interface as a plugin for Visual Studio Code to configure Camel integrations. This graphical approach significantly eases the process of route definition, as it avoids syntax and logical errors that may happen when manually writing a route in a text file. \emph{Karavan} supports all the components officially included in the Apache Camel Framework, that can be reused within a pipeline defined using the tool. A pipeline configured in Karavan can be automatically exported as a Java project for execution.
Additionally, Karavan supports the definition of custom Kamelets\footnote{https://camel.apache.org/camel-k/2.5.x/kamelets/kamelets.html}, which are reusable route templates designed to simplify route construction. 
To integrate this tool into the DataOps toolbox, we worked on enabling Chimera components from the interface. Moreover, we developed a set of reusable Kamelets specific to the project's needs and made them available through the Karavan catalogue.

\section{DataOps Pipelines Across Use Cases}\label{sec:usecases}

This section discusses the DataOps pipelines implemented for each use case~\cite{scrocca_dataops_2025} and how they address the challenges outlined in Section~\ref{sec:challenges}.

\textbf{Low-Code Edge Intelligence for Smart Factory}
The support for the description and discovery of OPC-UA devices through the Knowledge Graph Repository (KGR) is developed using the DataOps toolbox. More specifically, DataOps pipelines are used to insert and retrieve data from the KGR by carrying out two harmonization processes: a lifting operation that transforms data from an OPC UA NodeSet in XML format to RDF format according to the OPC UA ontology \cite{Schiekofer_2019}, and a lowering operation that queries data in RDF format and returns an OPC UA NodeSet in XML. The available operations are: (i) add an OPC UA NodeSet describing one or more devices, (ii) search for specific devices via SPARQL queries, (iii) retrieve the OPC UA XML representation of an entire NodeSet or of specific OPC UA nodes. Each operation is exposed as an HTTP endpoint leveraging a proper node connector to provide a comparable experience to the one defined for WoT discovery.
The use of the DataOps toolbox in this case highlights the potential for handling complex requirements in ensuring the interoperability of static nodes' information. First, we address the complexity involved in the mapping rules required for converting the OPC UA description of nodes to and from the chosen reference ontology. In the lifting, two subsequent mapping processes are required on the input NodeSet with the second one accessing as input both the input XML and the RDF result of the first mapping operation. EIP components provided by Camel support this scenario without requiring custom logic to be developed. Additionally, the second mapping process for the lifting requires a stateful computation supported by the \texttt{mapping-template} tool integrated in Chimera.
As a second aspect, the toolbox allows us to address the interface integration requirements. The Chimera components adopted for the mapping operations support: (i) the usage of both in-memory and remote triplestore for the storage and retrieval of RDF data, (ii) the usage of named graphs to efficiently handle the interdependencies between different OPC UA Nodesets. Additionally, the pipelines are easily configured with existing Camel components to expose an API for the storage and retrieval of devices' information. 
The DataOps pipeline successfully supports the import of the OPC UA Nodesets associated with all the industrial devices in the demonstration factory. The structured description of the devices in the KGR enables the definition of recipes involving both OPC UA and WoT devices and the retrieval of nodes required for their execution. 
The Mendix platform serves as a low-code solution for defining and orchestrating various recipes, each tailored to different target configurations of the product. Leveraging the DataOps pipeline, Mendix accesses structured data for both standards within the KGR through a unified API, streamlining application development based on the Recipe Model.
To further enhance usability, we introduced an LLM interface that bridges Mendix and the KGR. This allows users to retrieve relevant devices for specific tasks by simply entering natural language queries. The LLM translates these queries into SPARQL, which are then executed via the KGR API~\cite{nilay2024}.
This integrated pipeline—combining DataOps, the Mendix low-code platform, and the LLM—significantly simplifies the development of shopfloor solutions across heterogeneous industrial devices and standards.

\textbf{Smart Vehicle to Infrastructure}
In this use case, the DataOps toolbox is leveraged to implement a runtime solution for data stream fusion and harmonisation from different sensors and public transport data sources. The DataOps pipeline implemented is responsible for: (i) retrieving data from heterogeneous data sources, (ii) harmonising the data according to a common set of RDF vocabularies, (iii) implementing a fusion logic enriching the integrated output, (iv), generating and forwarding the output RDF stream to downstream components for further processing (e.g., leveraging a stream reasoning engine for computing traffic indicators). 
Dynamic data from the nodes (e.g., radar, LiDAR, camera, connected vehicle, etc.) are processed with AI-enabled capabilities and exposed via WebSockets or NATS~\cite{nats} in JSON format. Data sources adopt heterogeneous formats but contain similar and complementary information on the number, position, and type of vehicles detected.
Additionally, GTFS\footnote{GTFS Schedule and Realtime specifications available at \url{https://gtfs.org/}} data sources are considered to collect static (ZIP file over HTTP) and dynamic information (binary payload over HTTP) on public transport vehicles.
For each data source, a DataOps pipeline is configured with the appropriate node connector (according to the input protocol) and a mapping processor executing a set of declarative mapping rules from the input schema and data format to a shared RDF representation. Two ontologies are employed for this purpose: the ASAM OpenXOntology~\cite{asam}, which models road and vehicle-related data, and the SOSA ontology~\cite{Janowicz_2019}, which is used to describe sensor-generated data. Together, these ontologies provide a standardized and meaningful representation of traffic and sensor data, enabling data integration. Additionally, Wikidata identifiers are used to annotate the class of identified vehicles.
Once harmonized data from different data sources are integrated in a single RDF graph, a custom data fusion logic is applied to link nodes referring to potentially matching observations. For example, tram positions from the GTFS-RT stream help in correcting wrong observations from radar that may interpret a tram as one or more trucks. The generated output is pushed to a NATS server for further processing by other nodes. On the edge to support decision making for the traffic light behaviour, and in the cloud to support more intensive AI-based processing.
We experimented with different DataOps deployment templates by validating their applicability to embedded ARM devices on the edge. 
The latency introduced by the DataOps pipeline was measured together with memory and CPU usage to select the better deployment option\footnote{Test cases performed are reported in detail in~\cite{smartedged32}.}. For the target deployment, we managed to keep resource usage at a minimum (around 100MB memory consumption) while providing acceptable latency (below 100ms). The modular and configurable approach adopted by the DataOps toolbox facilitated the reuse and customisation of the generated pipeline for its deployment in different RSU at different intersections~\cite{grassi2025typhon}. Moreover, the declarative approach allowed iterative modification of the mapping rules (e.g., to change the target RDF output or the conditions for matching observations) directly by our stakeholders in Helsinki.

\textbf{Smart Health Rehabilitation}
The DataOps toolbox is applied in this use case to support different aspects of the presented CSI approach. A first DataOps pipeline supports the adoption of a custom low-code user interface (\emph{Semantic Recipe Interface}) to enable the therapist to define and assign exercises for a patient. The output of such a tool is processed by the pipeline to obtain an interoperable representation according to the Recipe Model and store it in the KGR.
A second DataOps pipeline supports the matching between a recipe and the devices available for the patient and described through the WoT TD in the KGR. Given a recipe identifier, the pipeline returns, for each capability defined by the recipe, the list of devices that could cover. The output is serialised in a JSON format that can be processed by the \emph{Recipe-Device Matcher}. Finally, a third DataOps pipeline supports the provision of additional data defined by the Recipe to the \emph{Recipe Orchestrator} by converting it to a target JSON format. As a result, the orchestrator performs the onboard of the available devices for the current exercise. Then, it applies AI-enabled algorithms to incoming data for monitoring the execution of the exercise by the patient. In this case, the components for the execution of the recipe are deployed on a tablet given to the patient that communicates with the required devices, while the DataOps pipelines are hosted in the cloud together with the KGR. 
This use case also effectively validated the use of the Karavan tool for defining the DataOps pipeline. Indeed, the drag-and-drop editor was successfully used to define the pipelines and configure them with the corresponding mapping rules. Finally, the pipelines were exported using the YAML DSL for execution via an appropriate deployment template. This approach also makes it possible to modify the YAML pipeline configuration and/or the mapping rules without a rebuild of the executable.

\section{Evaluation and Lessons Learned}\label{sec:validation}

This section discusses an evaluation of the proposed approach and the lessons learned.

\textbf{User evaluation.} The stakeholders involved in the validation appreciated the configurability and customisability of the DataOps toolbox that allowed them to tackle various issues to support data interoperability for their demonstrators. Additionally, we received positive feedback on the application of a declarative approach, both for the mapping rules and the pipeline specification, which facilitated their understanding in editing and adapting the delivered pipelines. One drawback they highlighted is that configuring the DataOps pipeline requires knowledge of the Apache Camel ecosystem, and the definition of mapping rules is primarily accessible to users with developer experience. Nevertheless, they evaluated this initial learning curve as acceptable, considering the broad applicability of the tool to support the implementation of various applications across devices.
Another aspect appreciated is related to the various ready-to-be-used deployment templates for DataOps pipelines that support both edge and cloud requirements. The users valued positively the possibility of deploying the same pipeline in different deployment environments with minimum effort and with scalability guarantees.

\textbf{Significance evaluation.} The proposed approach for CSI based on Semantic Web technologies is aligned with the well-known WoT specifications that are recognized by the research community and by the industry as a suitable solution to guarantee interoperability. The collaboration among devices is key for enabling distributed swarm applications and guaranteeing interoperability in the communication between devices is mandatory. The existing solution based on the WoT specification is not suitable for all the use cases and may not have enough expressiveness to support all devices (e.g., the OPC UA case mentioned in this paper). Additionally, the mediation between devices is not only a matter of protocols but may require very complicated data and schema transformations of payloads that can not be handled with a generalised approach. In this context, the proposed DataOps toolbox offers a modular and configurable solution to address these issues while adopting state-of-the-art approaches for semantic interoperability and declarative knowledge graph construction.

\textbf{Uptake and impact evaluation.} The implemented DataOps pipelines demonstrated that continuous semantic integration is achievable even across heterogeneous devices and complex, diverse standards spanning the cloud-edge continuum. Moreover, their usage in real-world deployments validated the applicability of the approach in production environments. The enabled interoperability of OPC-UA devices is considered a key asset for Siemens to develop new solutions for their customers and investigate further the usage of Mendix for low-code programmability of industrial machines. Furthermore, the availability of static and runtime information in a structured and interoperable format enables new scenarios leveraging LLMs for natural language interfaces. 
The pipelines for the smart traffic scenario have been positively evaluated by the stakeholders, who are planning further developments to take into account the innovation in communications with connected vehicles and the growing computational power in road infrastructure. Additionally, the interoperable data generated has also emerged as a potential source of open data that may interest the municipality or third-party developers.
Finally, the implementation of the approach for the rehabilitation use case has demonstrated its adaptability to address the heterogeneity of patient devices and its suitability for simplifying the setup of the tele-rehabilitation environment.  

\textbf{Lessons learned.} 
An important aspect that emerged in validating our technologies in these use cases is the need for stakeholders to gain expertise in Semantic Web technologies. To bridge this gap, we adopted low-code approaches for configuring DataOps pipelines, and we preliminarily investigated the usage of Large Language Models (LLMs) to further streamline the user experience. However, from the use cases, it emerges that mapping rules and integration logic can often be associated with complex requirements. In these cases, the DataOps toolbox represents a flexible solution to customise declarative pipelines by introducing specific customisations~\cite{grassi2025typhon}. 
Another key takeaway was the difficulty of finding standardised vocabularies across different domains. The lack of universally accepted vocabularies adds an additional step to implementing the proposed approach, possibly requiring the development of custom ontologies or extensions to address use case specificities.
Finally, we learned the importance of monitoring resource consumption closely. Balancing performance with sustainability is crucial, and we focused on enabling the instrumentation of DataOps pipelines\footnote{\url{https://github.com/cefriel/chimera-deployment-templates/tree/micrometer-stats}} to support their observability \cite{scrocca2020kaiju} and efficient resource management.

\section{Conclusions}\label{sec:conclusions}

In this paper, we discussed how Semantic Web technologies can effectively support the low-code programmability of AI applications leveraging the collaboration of multiple devices between edge and cloud. 

We presented an approach for Continuous Semantic Integration (CSI) of devices, and by eliciting challenges from practical use cases in various domains, we focused on the requirements for DataOps operations. A modular and flexible toolbox was proposed to address these aspects and support the implementation of DataOps pipelines for heterogeneous requirements. The toolbox provides a set of reusable and extensible components that can be configured via low-code approaches and easily deployed via pre-defined templates.
The adoption of the proposed solution in the three use cases demonstrated the toolbox's flexibility in addressing interoperability challenges and enabling the CSI approach. On one hand, we supported the discovery and matchmaking of nodes to execute specific applications; on the other hand, we enabled mediated data exchanges at runtime to support cooperation between nodes. We validated the pipelines in production environments and with various stakeholders acknowledging their effectiveness in supporting the proposed challenges.

As future work, we would like to further investigate the role of Large Language Models in supporting the proposed CSI approach and reducing the effort required by users in configuring applications. In particular, considering the DataOps toolbox, we will focus on minimising the effort required to select the required components and configure pipelines. Finally, we plan to work on the definition of reusable DataOps pipelines that could act as components compliant with the WoT architecture. 

\begin{small}
\paragraph*{Supplemental Material Statement:}
Public SmartEdge deliverables are available at \url{https://smart-edge.eu/deliverables}. The Chimera framework used as a basis for the DataOps toolbox is hosted at \url{https://github\-.com/cefriel/\-chimera} and the deployment templates at \url{https://\-github.com/cefriel/chimera-deployment-templates}. The DataOps pipelines developed for the SmartEdge use cases are confidential and can not be shared, but a set of diagrams providing additional details on their implementation is made available at \url{https://doi.org/10.5281/zenodo.15379925}. The SmartEdge Schema is available at \url{https://w3id.org/smartedge/smartedge-schema} and the Recipe Model at \url{https://w3id.org/smartedge/recipe-model}. 
\end{small}

\begin{credits} \subsubsection{\ackname} 
The presented research was partially supported by the SmartEdge project, funded under the Horizon Europe RIA Research and Innovation Programme (Grant Agreement 101092908). The authors would like to acknowledge Kirill Dorofeev (Siemens), Kari Koskinen and Mehrdad Bagheri (Conveqs), Davide Calvaresi and Berk Buzcu (HES-SO) for their support in the implementation of the use cases.

This preprint has not undergone peer review or any post-submission improvements or corrections. The Version of Record of this contribution will be published in The Semantic Web – ISWC 2025.
\end{credits}

%
% ---- Bibliography ----
%
% BibTeX users should specify bibliography style 'splncs04'.
% References will then be sorted and formatted in the correct style.
%
% \bibliographystyle{splncs04}
% \bibliography{mybibliography}
%
\bibliographystyle{splncs04}
\bibliography{biblio}

\end{document}